\documentclass[preprint,onecolumn,nofootinbib]{revtex4}
\pdfoutput=1

\usepackage[colorlinks=true,linkcolor=blue,urlcolor=blue,filecolor=black,citecolor=red,pdfstartview=FitV,pdftitle={},pdfsubject={},pdfkeywords={},pdfpagemode=None,bookmarksopen=true]{hyperref}
\usepackage{graphicx}
\usepackage{amsmath}
\usepackage{amsfonts}
\usepackage{amssymb,ulem}
\usepackage{color}%
\usepackage{multirow}
\usepackage{dcolumn}
\newcommand{\kk}{\big\rangle}
\newcommand{\bb}{\big\langle}

\newcommand{\f}{\begin{equation}}
\newcommand{\ff}{\end{equation}}
\newcommand{\fa}{\begin{eqnarray}}
\newcommand{\ffa}{\end{eqnarray}}

\begin{document}
\title{Holograhic two-currents model with coupling and its conductivites}
\author{Dan Zhang $^{1}$}
\thanks{danzhanglnk@163.com}
\author{Zhenhua Zhou $^{2}$}
\thanks{dtplanck@163.com}
\author{Guoyang Fu $^{3}$}
\thanks{FuguoyangEDU@163.com}
\author{Jian-Pin Wu $^{3}$}
\thanks{jianpinwu@yzu.edu.cn}
\affiliation{$^1$~College of Physical Science and Technology, Bohai University, Jinzhou 121013, China}
\affiliation{$^2$~School of Physics and Electronic Information, Yunnan Normal University, Kunming, 650500, China}
\affiliation{$^3$~Center for Gravitation and Cosmology, College of Physical Science and Technology, Yangzhou University, Yangzhou 225009, China}

\begin{abstract}

We implement a holographic gravity model of two gauge fields with a coupling between them, which is dual to a two-currents model. An analytical black brane solution is obtained. In particular, we work out the expressions of conductivities with coupling and find that the expressions of conductivities are directly related to the coupling parameter $\theta$. It is the main topic of our present work. As an application, we calculate the conductivities by the scheme outlined here and briefly discuss the properties of the conductivities. An interesting property is that as the coupling $\theta$ increases, the dip at low frequency in $Re[\sigma_A]$/$Re[\sigma_B]$ becomes deepening and then turns into a hard-gap-like when $\theta=1$, which is independent of the doping $\chi$. Some monotonic behaviors of the conductivities are also discussed.

\end{abstract}
\maketitle

\newpage
\section{Introduction}

AdS/CFT (Anti-de-Sitter/Conformal field theory) correspondence, also referring to holography, provides us a way to study the dynamics of certain strongly-coupled condensed matter systems \cite{Maldacena:1997re,Gubser:1998bc,Witten:1998qj,Aharony:1999ti}.  In the so-called bottom-up approaches, we can study some universal properties of the dual system by constructing a simple gravitational model. Along this direction, some interesting holographic models, for example, the holographic superconductor \cite{Hartnoll:2008vx}, holographic metal insulator phase transition (MIT) \cite{Donos:2012js,Ling:2014saa}
and (non-) Fermi liquid \cite{Liu:2009dm}, have been implemented. These studies provide some physical insight into the associated mechanisms of the strongly-coupled systems and the universality class of them.

Recently, the holographic two-currents models are increasingly drawn attention, see \cite{Kiritsis:2015hoa,Bigazzi:2011ak,Huang:2020iyw,Iqbal:2010eh,Baggioli:2015dwa,Rogatko:2017tae,Rogatko:2020vtz,Seo:2016vks} and references therein. In these models, a pair of $U(1)$ gauge fields $A$ and $B$ in bulk are introduced\footnote{Some holographic models with two gauge fields are also studied in \cite{Ling:2015exa,Ling:2020mwm,Ling:2017naw,Ling:2016wyr,Tarrio:2011de,Alishahiha:2012qu}, in which only one of gauge fields is treated as the real Maxwell field and we only concentrate on its transport properties. In \cite{Ling:2015exa}, another gauge field is treated as an auxiliary field, which is introduced to obtain an insulating phase with a hard gap. On top of this novel holographic insulator constructed in \cite{Ling:2015exa}, a holographic superconductor is also built \cite{Ling:2017naw}. Further, \cite{Ling:2020mwm} also introduces a coupling between these two gauge fields and studies the superconducting instability. In \cite{Ling:2016wyr}, another gauge field is introduced to induce metal-insulator phase transition over Gubser-Rocha background \cite{Gubser:2009qt} in the limit of of zero temperature. While in \cite{Tarrio:2011de,Alishahiha:2012qu}, the additional gauge field also plays the role of the auxiliary field to implement charged hyperscaling or Lifshitz black hole background. Therefore, these models are still treated as single current models.}. Therefore, we have two independently conserved currents, which relate to different kinds of chemical potentials or charged densities in the dual boundary field theory. The mismatch of the two controllable chemical potentials or charge densities induces the unbalance of numbers. In \cite{Kiritsis:2015hoa,Baggioli:2015dwa,Huang:2020iyw}, the ration of the two chemical potentials is proposed to simulate the effect of doping. In \cite{Bigazzi:2011ak}, they propose the holographic two-currents model has a counterpart of Mott's two-currents model \cite{Mott:1936v1,Mott:1936v2}. The chemical potential mismatch is interpreted as a chemical potential for a $U(1)_B$ ``spin'' symmetry \cite{Bigazzi:2011ak,Iqbal:2010eh}. And then, on top of the two-current model, they construct an unbalanced s-wave superconductor by introducing a charged complex scalar field \cite{Bigazzi:2011ak}, which is a simple extension of holographic superconductor in \cite{Hartnoll:2008vx}. Also they study ``charge'' and ``spin'' transport \cite{Bigazzi:2011ak}\footnote{Some related works are also explored, see for example \cite{Erdmenger:2011hp,Dutta:2013osl,Correa:2019ivh,Musso:2013ija,Alsup:2012kr,Hafshejani:2018svs}.}. In \cite{Rogatko:2017tae,Rogatko:2020vtz,Seo:2016vks}, the holographic two-currents model is used to describe the nature of graphene.

Most of these works do not contain the coupling between two gauge fields.
Recently, there have been a small number of works beginning to concentrate on the effect of the coupling between two gauge fields \cite{Baggioli:2015dwa,Rogatko:2017tae,Rogatko:2020vtz}. The coupling between two gauge fields provides additional degree of freedom in the dual boundary field theory.
In \cite{Baggioli:2015dwa}, they build a holographic superconductor model containing a non-trivial higher derivative therm of axionic field breaking translational symmetry, a complex scalar field breaking $U(1)$ symmetry and two $U(1)$ gauge fields. In particular, the complex scalar field provides a non-trivial coupling between two gauge fields. In this way, they implement a superconducting dome-shaped region on the temperature-doping phase diagram. But the study of the transport properties of this model is absent. In \cite{Rogatko:2017tae}, they introduce a simple coupling term between two gauge fields and explored its thermoelectric transport properties of its holographic dual boundary field theory describing graphene. In \cite{Rogatko:2020vtz}, they show that there is a bound on the conductivity depending on the coupling between both gauge fields. At the same time, their study also indicates that even strong disorder cannot still induce a MIT in holographic two-currents model as that with single current \cite{Grozdanov:2015qia}.

In this paper, we shall construct a holographic two-currents model with coupling. We derive the expression of the frequency dependent thermoelectric transport and explore its properties. Especially, we concentrate on the effect of the coupling between two gauge fields. Our paper is organized as what follows. In Sec.\ref{sec-framework}, we describe the holographic framework of the two-currents model with coupling and work out the analytical double charged RN-AdS black brane solution. In Sec.\ref{sec-hcon}, by standard holographic renormalized procedure, we obtain the expressions of the holographic conductivities for our two-currents model with coupling. And then, the properties of the conductivities are discussed in Sec.\ref{sec-num}. The conclusions and discussions are presented in Sec.\ref{sec-con-dis}.

\section{The double charged RN-AdS black brane}\label{sec-framework}

The action of the gravity dual for a simple two-currents model with coupling is
\begin{eqnarray}
\label{action}
&&S=\int d^{4}x \sqrt{-g}\,
\left(R+\frac{6}{L^2}
+\mathcal{L}_M
\right)\,,\nonumber \\
&&\mathcal{L}_M=-\frac{1}{4}F_{\mu\nu}F^{\mu\nu}-\frac{1}{4}G_{\mu\nu}G^{\mu\nu}-\frac{\theta}{2}F_{\mu\nu}G^{\mu\nu}\,,  
\end{eqnarray}
where we set the AdS radius $L=1$ in this paper. $F=dA$ and $G=dB$ are the field strengths of the two gauge fields $A$ and $B$, respectively. Above we bring in a coupling term between two gauge fields and $\theta$ denotes the coupling strength. Notice that when $\theta^2=1$, we can combine both gauge fields $A$ and $B$ into a new gauge field
such that both gauge fields are indistinguishable.

From the action \eqref{action}, we derive the equations of motion (EOMs) as what follows
\begin{subequations}
\label{eom}
\begin{align}
&R_{\mu\nu}-\frac{1}{2}R g_{\mu\nu}-\frac{3}{L^2}g_{\mu\nu}-\frac{1}{2}T^{(A)}_{\mu\nu}-\frac{1}{2}T^{(B)}_{\mu\nu}-\theta T^{(AB)}_{\mu\nu}=0\,,
\label{EE}\\
&\nabla_{\mu}F^{\mu\nu}=0 \,,\label{MaxwellA} \\
&\nabla_{\mu}G^{\mu\nu}=0\,.\label{MaxwellB}
\end{align}
\end{subequations}
The last three terms in Einstein equation \eqref{EE} are defined as
\begin{subequations}
\label{EMTensor}
\begin{align}
&T^{(A)}_{\mu\nu}=F_{\mu\rho}F_{\nu}^{\ \rho}-\frac{1}{4}g_{\mu\nu}F^2\,,
\
\\
&T^{(B)}_{\mu\nu}=G_{\mu\rho}G_{\nu}^{\ \rho}-\frac{1}{4}g_{\mu\nu}G^2\,,
\
\\
&T^{(AB)}_{\mu\nu}=F_{(\mu|\rho|}G_{\nu)}^{\ \rho}-\frac{1}{4}g_{\mu\nu}F_{\alpha\beta}G^{\alpha\beta}\,,
\end{align}
\end{subequations}
where the symmetry bracket means $A_{(\mu\nu)}=(A_{\mu\nu}+A_{\nu\mu})/2$.

One can obtain the following double charged RN-AdS black blane from the theory \eqref{action}
\begin{subequations}
\label{sol}
\begin{align}
&
\label{ds}
ds^2=\frac{1}{u^2}\Big(-f(u) dt^2+\frac{du^2}{f(u)}+dx^2+dy^2\Big)\,,
\
\\
&
\label{fr}
f(u)=1-\Big(1+\frac{u^4_+(q_A^2+q_B^2+2\theta q_A q_B)}{4}\Big)\Big(\frac{u}{u_+}\Big)^3+\frac{u^4_+(q_A^2+q_B^2+2\theta q_A q_B)}{4}\Big(\frac{u}{u_+}\Big)^4\,,
\
\\
&
\label{Atr}
A_t(u)=\mu-q_A\,u\,,~~~~~~B_t(u)=\delta\mu-q_B\,u\,,
\end{align}
\end{subequations}
where the AdS boundary and the horizon are located at $u=0$ and $u=u_+$, respectively. $\mu$ and $\delta\mu$ are the chemical potentials of the gauge fields $A$ and $B$ in the dual boundary field theory. $q_A$ and $q_B$ are two integration constants relating to the chemical potentials by $q_A=\mu/u_+$ and $q_B=\delta\mu/u_+$, which are determined by the horizon conditions $A_t$ and $B_t$ satisfied. Therefore, there are two controllable chemical potentials causing the unbalance of numbers. The ration $\chi=\frac{\delta\mu}{\mu}$ stands for the strength of the unbalance. In \cite{Kiritsis:2015hoa,Baggioli:2015dwa,Huang:2020iyw}, it was used to simulate the doping.
The coupling strength $\theta$ characters the charged impurities coupling strength in the dual field theory. When $\theta=0$, the black brane solution \eqref{sol} reduces to that studied in \cite{Bigazzi:2011ak}.

According to the solution \eqref{sol}, the Hawking temperature can be straightforward calculated as
\begin{align}
\label{tem}
T=-\frac{f'(u_+)}{4\pi}=\frac{1}{4\pi}\Big(\frac{3}{u_+}-\frac{u_+(\mu^2+\delta\mu^2+2\theta\mu\,\delta\mu)}{4}\Big)\,.
\end{align}
It corresponds to the temperature of the dual field theory.

To obtain the other thermodynamical quantities, we write down the renormalized action following the strategy in \cite{Caldarelli:2016nni}
\begin{align}
S_{ren}=S+\int_{u=\varepsilon} d^3x \sqrt{-\gamma}\,\Big(2K-\frac{4}{L}-LR[\gamma]\Big)\,.
\end{align}
$\gamma$ is the determinant of the boundary induced metric $\gamma_{\mu\nu}=g_{\mu\nu}-n_\mu n_\nu$ and $K$ is the trace of the extrinsic curvature $K_{ij}=-\frac{1}{2 L}u\partial_u\gamma_{ij}$. They take value at the UV cut-off $u=\varepsilon$ and then is sent to zero following the holographic renormalized procedure.
Notice that $n_{\mu}$ is the out-point normal vector of the UV cut-off surface.

And then, one obtains the corresponding on-shell renormalized action, which reads
\begin{align}
S^{os}_{ren}=\int d^{3}x \Big[\frac{1}{u_+^3}+\frac{\mu^2+\delta\mu^2+2\theta\mu\,\delta\mu}{4u_+}\Big]\,.
\end{align}
Immediately, according to holography, the thermal potential $\Omega$ is worked out as
\begin{align}
\Omega=-V_2\Big(\frac{1}{u_+^3}+\frac{\mu^2+\delta\mu^2+2\theta\mu\,\delta\mu}{4u_+}\Big)\,,
\end{align}
where $V_2\equiv\int dxdy$.
Once we have the thermal potential, it is easy to calculate the entropy density $s$, charge densities $\rho_A$ and $\rho_B$, which are
\begin{align}
s=-\frac{1}{V_2}\frac{\partial\Omega}{\partial T}=\frac{4\pi}{u_+^2}\,,\quad \rho_A=-\frac{1}{V_2}\frac{\partial\Omega}{\partial \mu}=\frac{\mu+\theta \delta\mu}{u_+}\,,\quad
\rho_B=-\frac{1}{V_2}\frac{\partial\Omega}{\partial\, \delta\mu}=\frac{\theta \mu+\delta\mu}{u_+}\,.
\end{align}
Also, the press and the energy density of the system can be given by
\begin{align}
&p=\Big(\frac{1}{u_+^3}+\frac{\mu^2+\delta\mu^2+2\theta\mu\,\delta\mu}{4u_+}\Big)\,,\\
&\epsilon=2\Big(\frac{1}{u_+^3}+\frac{\mu^2+\delta\mu^2+2\theta\mu\,\delta\mu}{4u_+}\Big)\,.
\end{align}
Note that the positive definiteness of charge densities $\rho_A$ and $\rho_B$ requires
\begin{eqnarray} \label{theta-region}
\begin{cases} \theta\geq-\chi\,, & \chi\leq 1 \cr
            \theta\geq-\frac{1}{\chi}\,, & \chi\geq 1
            \end{cases}\,.
\end{eqnarray}

\section{Holographic expression of the conductivities}\label{sec-hcon}

In \cite{Bigazzi:2011ak}, the holographic expression of the conductivities have been derived for the two-currents model without coupling. In the presence of a coupling between the two gauge fields, the case becomes subtle and we present the detailed derivation for the conductivities in this section.

Thanks to the rotational invariance in $x-y$ plane, we only need consider the conductivities along $x$-direction.
We first describe the conductivity matrix of the holographic two-currents model.
We denote the currents, external fields and conductivities associated to the gauge fields $A$ and $B$ as
$(J_A,E_A,\sigma_A)$ and $(J_B,E_B,\sigma_B)$, respectively.
At the same time, the external field $E_A$ also leads to the current $J_B$.
The associated conductivity is denoted as $\bar{\gamma}_{AB}$.
This process is reciprocal. $E_B$ also generates the current $J_A$ giving the associated conductivity $\bar{\gamma}_{BA}$. The time-reversal invariance results in $\bar{\gamma}_{AB}=\bar{\gamma}_{BA}\equiv\bar{\gamma}$.
In Refs.\cite{Mott:1936v1,Mott:1936v2,Fert:1968,Son:1987,Johnson:1987,Bigazzi:2011ak}, $\sigma_A$ and $\sigma_B$ are interpreted as electric conductivity and spin-spin conductivity, and correspondingly $\bar{\gamma}$ is the spin conductivity.
Both currents also induce some momentum, acting on the momentum operator $T_{tx}$ as source.
Besides, the temperature gradient leads to the heat current $Q=T_{tx}-\mu J_A-\delta\mu J_B$, which induces thermal conductivity $\bar{\kappa}$.
The another two transport quantities are the thermo-electric and thermo-spin conductivities associated to the transport of heat. We denote them as $\alpha$ and $\beta$.
And then, Ohm's law can be expressed as
\begin{align}
\left(
  \begin{array}{c}
    J_A \\
    Q \\
    J_B \\
  \end{array}
\right)=
\left(
  \begin{array}{ccc}
    \sigma_A & \alpha T & \bar{\gamma} \\
    \alpha T & \bar{\kappa} T & \beta T \\
    \bar{\gamma} & \beta T & \sigma_B\\
  \end{array}
\right)
\left(
  \begin{array}{c}
    E_A \\
    -\nabla T/T \\
    E_B \\
  \end{array}
\right)\,.
\label{ohm}
\end{align}
The conductivity matrix is symmetric because of the time-reversal symmetry.

Now, we turn to derive the expressions of the conductivities in our holographic two-currents model.
To this end, we turn on the bulk fluctuations $A_x$, $B_x$ and $g_{tx}$, which provide the source for the currents $J^x_A$ and $J^x_B$,
the stress energy tensor component $T^{tx}$ in the dual boundary field theory.
Explicitly, we set
\begin{subequations}
\label{ansatzpur1}
\begin{align}
\label{metricp}
&ds^2=\frac{1}{u^2}\Big(-f(u)dt^2+\frac{du^2}{f(u)}+\delta_{ij}dx^idx^j+2g_{tx}(u,t)dtdx\Big)\,,
\
\\
&
A=A_t(u)dt+A_x(u,t)dx\,,\quad B=B_t(u)dt+B_x(u,t)dx\,.
\end{align}
\end{subequations}
And then, taking a simple time dependence for the fluctuations as
\fa
g_{tx}(u,t)\equiv \frac{1}{u^2} h_{tx}(u)e^{-i\omega t}\,,\quad A_x(u,t)\equiv A_x(u)e^{-i\omega t}\,,\quad B_x(u,t)\equiv B_x(u)e^{-i\omega t}\,,
\ffa
one obtains the linear EOMs for $h_{tx},A_x,B_x$ as
\footnote{One can check that the two non-vanished Einstein equations along $tx$ and $zx$ directions are equivalent, thus we just list one of them here.}
\begin{subequations}
\label{EOM-perturbation}
\begin{align}
&h'_{tx}=u^2\Big((q_A+\theta q_B)A_x+(q_B+\theta q_A)B_x\Big)\,,\\
&(fA'_x)'+\frac{\omega^2  A_x}{f}-q_A h'_{tx}=0\,,\\
& (f B'_x)'+\frac{\omega^2  B_x}{f}-q_B h'_{tx}=0\,.
\end{align}
\end{subequations}
It is easy to see that among the above three EOMs, only two of them are independent.
In the limit of $u\rightarrow 0$,  the fields follow
\begin{subequations}
\label{asymbdy1}
\begin{align}
&A_x=A_x^{(0)}+A_x^{(1)}u+\cdots\,,
\
\\
&B_x=B_x^{(0)}+B_x^{(1)}u+\cdots\,,
\
\\
&h_{tx}=h_{tx}^{(0)}+h_{tx}^{(1)}u^3+\cdots\,.
\end{align}
\end{subequations}

To have a well-defined bulk variational problem, we write down the following renormalized action
\begin{eqnarray}
S_{ren}=S+\int_{u=\varepsilon} d^3x \sqrt{-\gamma}\,(2K-4)\,.
\end{eqnarray}
Making the variation of the on-shell action, one has
\begin{align}
\delta S^{on-shell}_{ren}&= \int_{u=\varepsilon} d^3x \sqrt{|\gamma|}\Big(-K^{\mu\nu}+K\gamma^{\mu\nu}-2\gamma^{\mu\nu}\Big)\delta \gamma_{\mu\nu}
-\int_{u=\varepsilon} d^3x\sqrt{|\gamma|}n_\mu( F^{\mu\nu}+\theta G^{\mu\nu})\delta A_\nu
\nonumber\\
&-\int_{u=\varepsilon} d^3x\sqrt{|\gamma|}n_\mu( G^{\mu\nu}+\theta F^{\mu\nu})\delta B_\nu\,.
\end{align}
Further using the ansatz \eqref{ansatzpur1} and the UV expansion \eqref{asymbdy1}, we can evaluate the above equation as
\begin{align}
&\delta S^{on-shell}_{ren}=\int_{u=\varepsilon} d^3x \Big[\frac{2}{u^3}(1-\frac{1}{\sqrt{f}})h_{tx}-\frac{1}{2u^2}h'_{tx}\Big]e^{-i\omega t}\delta h_{tx}e^{-i\omega t}\nonumber\\
&+\int_{u=\varepsilon} d^3x \Big[\big(f(A'_x+\theta B'_x)-(q_A+\theta q_B) h_{tx}\big)e^{-i\omega t}\delta A_xe^{-i\omega t}\nonumber\\
&+ \big(f(\theta A'_x+B'_x)-(\theta q_A+q_B) h_{tx}\big)e^{-i\omega t}\delta B_xe^{-i\omega t}\Big]\nonumber\\
&=\int_{u=\varepsilon} d^3x \Big[\frac{2}{u^3}(1-\frac{1}{\sqrt{f}})h^{(0)}_{tx}-\frac{1}{2}\big((q_A+\theta q_B)A^{(0)}_x+(\theta q_A+q_B)B^{(0)}_x\big )\Big]e^{-i\omega t}\delta h^{(0)}_{tx}e^{-i\omega t}\nonumber\\
&+\int_{u=\varepsilon} d^3x \Big[\big(A^{(1)}_x+\theta B^{(1)}_x -(q_A+\theta q_B) h^{(0)}_{tx}\big)e^{-i\omega t}\delta A^{(0)}_xe^{-i\omega t}\nonumber\\
&+ \big(\theta A^{(1)}_x+B^{(1)}_x-(\theta q_A+q_B) h^{(0)}_{tx}\big)e^{-i\omega t}\delta B^{(0)}_xe^{-i\omega t}\Big]\,.
\end{align}
According to the holographic dictionary
\begin{align}
\bb T^{\mu\nu}\kk(t)=2\frac{\delta S_{ren}^{on-shell}[g_{\mu\nu}^{(0)}]}{\delta g_{\mu\nu}^{(0)}(t)}\,,\qquad \bb J^\mu\kk(t)=\frac{\delta S_{ren}^{on-shell}[A_\mu^{(0)}]}{\delta A_\mu^{(0)}(t)}\,,
\end{align}
one obtains the expectation values of $J^x_A,\,J^x_B$ and $T^{tx}$ as
\begin{subequations}
\begin{align}
&\bb T^{tx}\kk(\omega)=\frac{4}{u^3}(1-\frac{1}{\sqrt{f}})h^{(0)}_{tx}-\Big((q_A+\theta q_B)A^{(0)}_x+(\theta q_A+q_B)B^{(0)}_x\Big)\,,\\
&\bb J^x_A\kk(\omega)=A^{(1)}_x+\theta B^{(1)}_x -(q_A+\theta q_B) h^{(0)}_{tx}\,,\\
&\bb J^x_B\kk(\omega)=\theta A^{(1)}_x+B^{(1)}_x-(\theta q_A+q_B) h^{(0)}_{tx}\,.
\end{align}
\end{subequations}
This source-response relation can be wrote in the matrix form,
\begin{align}
\left(
  \begin{array}{c}
    J^x_A(\omega) \\
    T^{tx}(\omega) \\
    J^x_B(\omega) \\
  \end{array}
\right)=
\left(
  \begin{array}{ccc}
     \dfrac{\delta A^{(1)}_x}{\delta A^{(0)}_x}+\theta\dfrac{\delta B^{(1)}_x}{\delta A^{(0)}_x}, & -(q_A+\theta q_B), &  \dfrac{\delta A^{(1)}_x}{\delta B^{(0)}_x}+\theta \dfrac{\delta B^{(1)}_x}{\delta B^{(0)}_x} \\
    -(q_A+\theta q_B), &  -\epsilon, & -(\theta q_A+q_B) \\
     \theta \dfrac{\delta A^{(1)}_x}{\delta A^{(0)}_x}+\dfrac{\delta B^{(1)}_x}{\delta A^{(0)}_x}, & -(\theta q_A+q_B),  &  \theta \dfrac{\delta A^{(1)}_x}{\delta B^{(0)}_x}+\dfrac{\delta B^{(1)}_x}{\delta B^{(0)}_x}\\
  \end{array}
\right)
\left(
  \begin{array}{c}
    A^{(0)}_x(\omega) \\
    h_{tx}^{(0)}(\omega) \\
   B^{(0)}_x(\omega) \\
  \end{array}
\right)\,,
\label{s-r1}
\end{align}
where the energy density is $\epsilon=-lim_{u\rightarrow0} 4 u^{-3}(1-f^{-1/2})$.
And then, we have the relation of the heat current $Q^x$, electric fields $E_{Ax},E_{Bx}$, and temperature gradient $\nabla_xT$ on the source, which are given by
\begin{subequations}
\begin{align}
&
\left(
  \begin{array}{c}
    J^x_A \\
    Q^{tx} \\
    J^x_B \\
  \end{array}
\right)=
\left(
  \begin{array}{ccc}
    1 & 0 & 0 \\
    -\mu &  1 & -\delta\mu \\
    0 & 0 & 1\\
  \end{array}
\right)
\left(
  \begin{array}{c}
    J^x_A \\
    T^{tx} \\
    J^x_B \\
  \end{array}
\right)\,,
\
\\
&
\left(
  \begin{array}{c}
    A^{(0)}_x \\
    h_{tx}^{(0)}\\
   B^{(0)}_x \\
  \end{array}
\right)=\frac{1}{i\omega}
\left(
  \begin{array}{ccc}
    1 & -\mu & 0 \\
    0 &  1 &  0 \\
    0 & -\delta\mu & 1\\
  \end{array}
\right)
\left(
  \begin{array}{c}
    E_{Ax} \\
    -\nabla_x T/T \\
    E_{Bx} \\
  \end{array}
  \right)\,.
\end{align}
\end{subequations}
Together with \eqref{s-r1} and comparing with Ohm's law \eqref{ohm},
one obtains the holographic expressions of alternating current (AC) conductivities of the two-currents model
\begin{subequations}
\label{exp-con}
\begin{align}
&\sigma_A=\frac{1}{i\omega}\Big(\dfrac{\delta A^{(1)}_x}{\delta A^{(0)}_x}+\theta\dfrac{\delta B^{(1)}_x}{\delta A^{(0)}_x}\Big)\,,
\
\\
&\sigma_B=\frac{1}{i\omega}\Big(\theta \dfrac{\delta A^{(1)}_x}{\delta B^{(0)}_x}+\dfrac{\delta B^{(1)}_x}{\delta B^{(0)}_x}\Big)\,,
\
\\
&\bar{\gamma}=\frac{1}{i\omega}\Big(\dfrac{\delta A^{(1)}_x}{\delta B^{(0)}_x}+\theta \dfrac{\delta B^{(1)}_x}{\delta B^{(0)}_x}\Big)
=\frac{1}{i\omega}\Big(\theta \dfrac{\delta A^{(1)}_x}{\delta A^{(0)}_x}+\dfrac{\delta B^{(1)}_x}{\delta A^{(0)}_x}\Big)\,,
\
\\
&\alpha T=-\mu\sigma_A-\delta\mu \,\gamma-\frac{q_A+\theta q_B}{i\omega}\,,
\
\\
&\beta T=-\delta\mu\sigma_B-\mu\,\gamma-\frac{\theta q_A+q_B}{i\omega}\,,
\
\\
&\bar{\kappa} T=\frac{1}{i\omega}\Big(-\epsilon-p+2\mu (q_A+\theta q_B)+2\delta\mu (\theta q_A+q_B)\Big)+ \mu^2\sigma_A+(\delta\mu)^2\sigma_B+2\mu \delta\mu\, \gamma
\,.
\end{align}
\end{subequations}
Since the two gauge fields are directly coupled in the gravity action, the conductivities $\sigma_A$ and $\sigma_B$ are also directly related to the coupling strength $\theta$.
As a result, the other conductivities $\bar{\gamma}$, $\alpha$ and $\beta$ are also related to the coupling $\theta$.

Until now, we have worked out the expressions of AC conductivities of the two-currents model in the presence of coupling, which is the main topic of our present paper. Using the expressions, we can explore the transport properties of holographic two-currents model with couple, for example, the superconductivity. As a simple application, here we calculate the AC conductivities of our present model and briefly discuss its properties.

\section{Numerical Results}\label{sec-num}

By numerically solving the EOMs \eqref{EOM-perturbation}, we can study the transport properties.
In the numerical calculation, by rescaling, the horizon location can be set as unity, i.e., $u_+=1$.
Thanks to the scaling symmetry, we take the chemical potential $\mu$ of gauge field $A$ as scaling unite.
Therefore, our theory is specified by the two dimensionless parameters $\hat{T}\equiv\frac{T}{\mu}$ as well as $\chi$, and the coupling parameter $\theta$.

\begin{figure}
\center{
\includegraphics[scale=0.8]{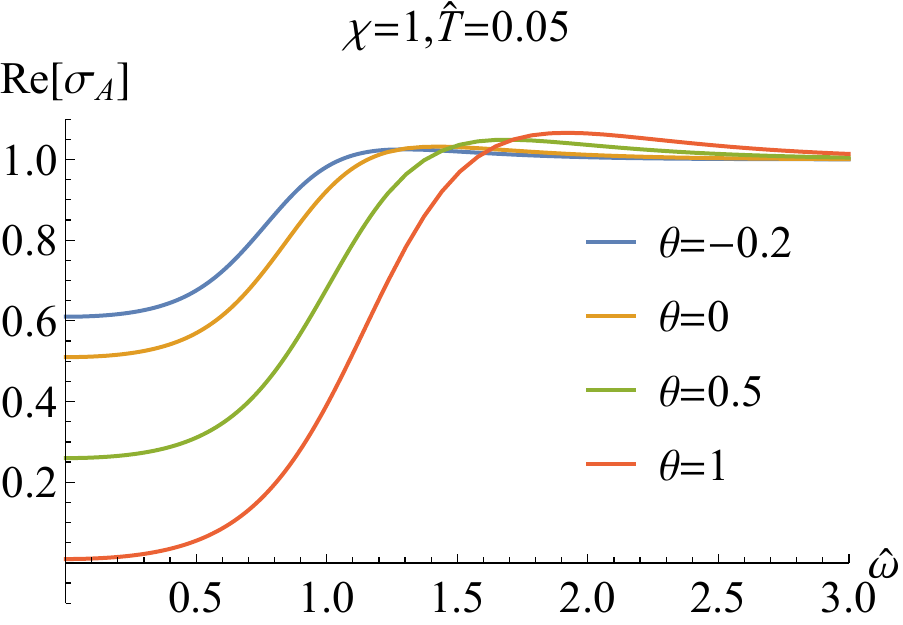}\ \hspace{0.5cm}
\includegraphics[scale=0.8]{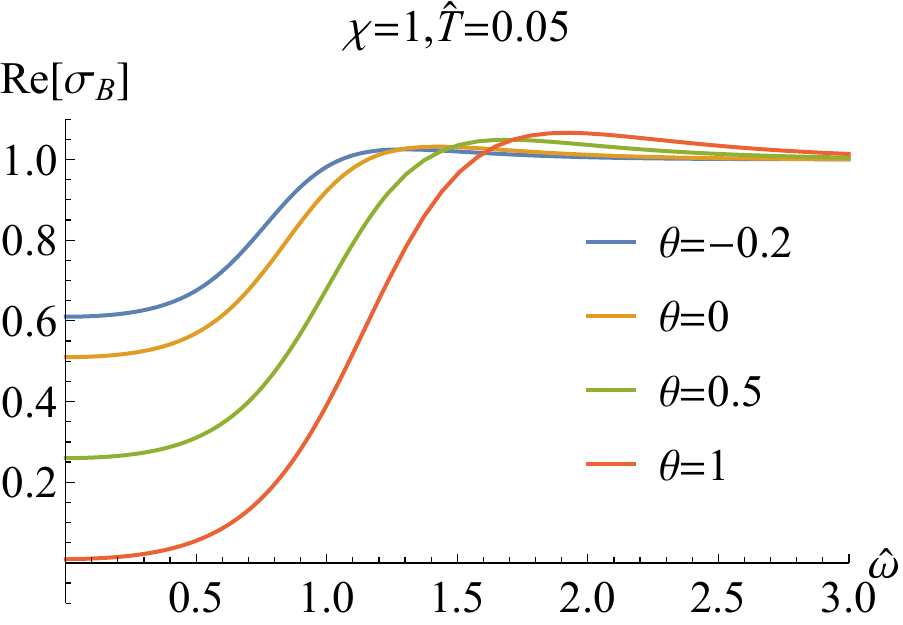}\ \\
\caption{\label{fig_sigmaAB_Chiv1} The real part of the conductivities $\sigma_A$ (left) and $\sigma_B$ (right) at $\hat{T}=0.05$ for $\chi=1$ but for different parameter $\theta$.
}}
\end{figure}
\begin{figure}
\center{
\includegraphics[scale=0.8]{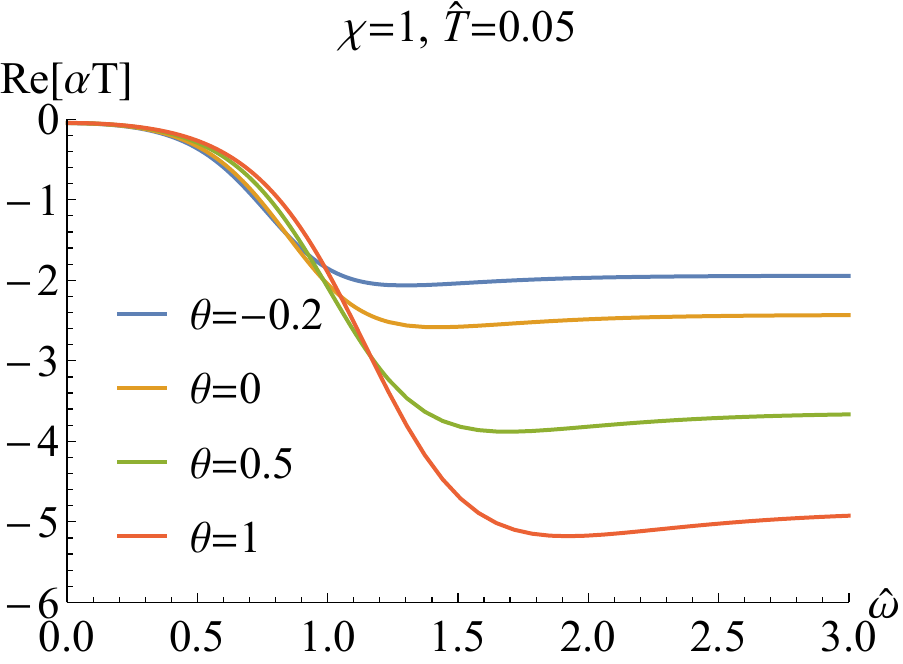}\ \hspace{0.5cm}
\includegraphics[scale=0.8]{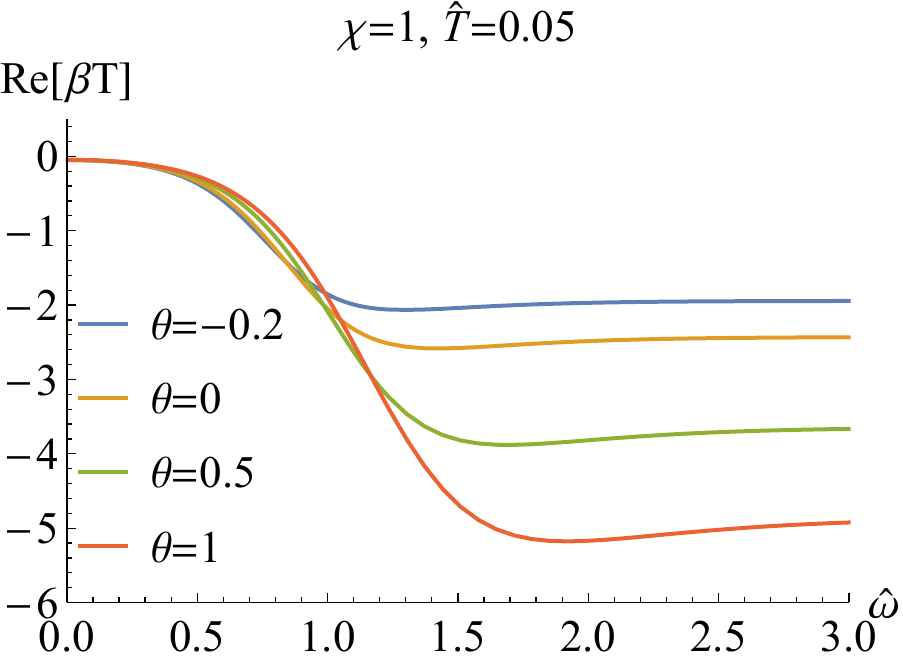}\ \\
\caption{\label{fig_alpha_Chiv1} The real part of the conductivities $\alpha T$ (left) and $\beta T$ (right) at $\hat{T}=0.05$ for $\chi=1$ but for different parameter $\theta$.
}}
\end{figure}

This holographic system possesses the following symmetries
\begin{subequations}
\label{symmetry}
\begin{align}
&
\sigma_A(\mu,\delta\mu,\theta,\hat{\omega})=\sigma_B(\delta\mu,\mu,\theta,\hat{\omega})\,,
\
\\
&
\alpha T(\mu,\delta\mu,\theta,\hat{\omega})=\beta T(\delta\mu,\mu,\theta,\hat{\omega})\,,
\end{align}
\end{subequations}
which can easily deduced from the EOMs \eqref{eom} and the expressions of the conductivity \eqref{exp-con}.
Numerically, we have also confirmed the above result.
It is similar to that without coupling $\theta$ \cite{Bigazzi:2011ak}. Especially, for $\mu=\delta\mu$, i.e., $\chi=1$, one has $\sigma_A=\sigma_B$ and $\alpha T=\beta T$, which are clearly seen in FIG.\ref{fig_sigmaAB_Chiv1} and FIG.\ref{fig_alpha_Chiv1}.

\begin{figure}
\center{
\includegraphics[scale=0.8]{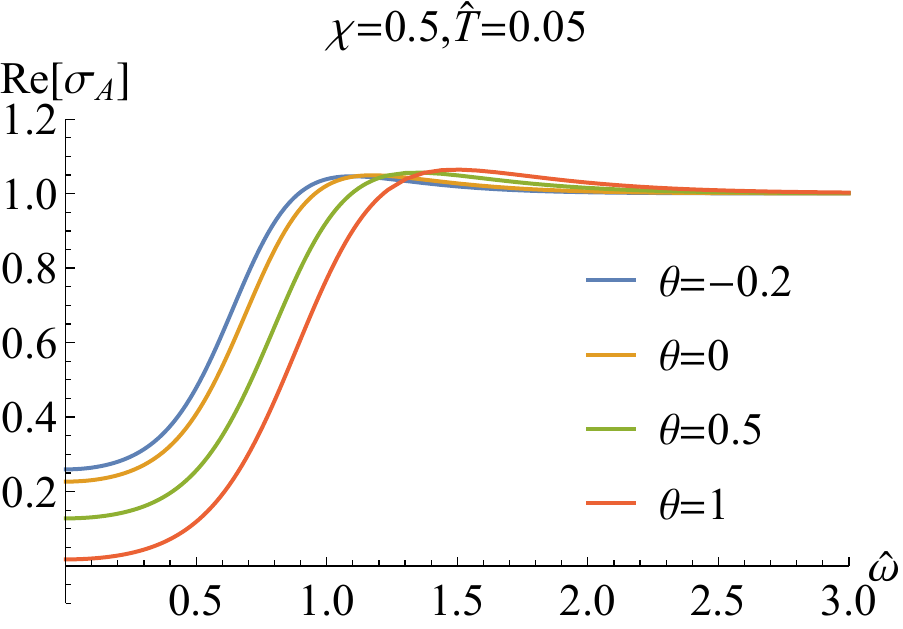}\ \hspace{0.5cm}
\includegraphics[scale=0.8]{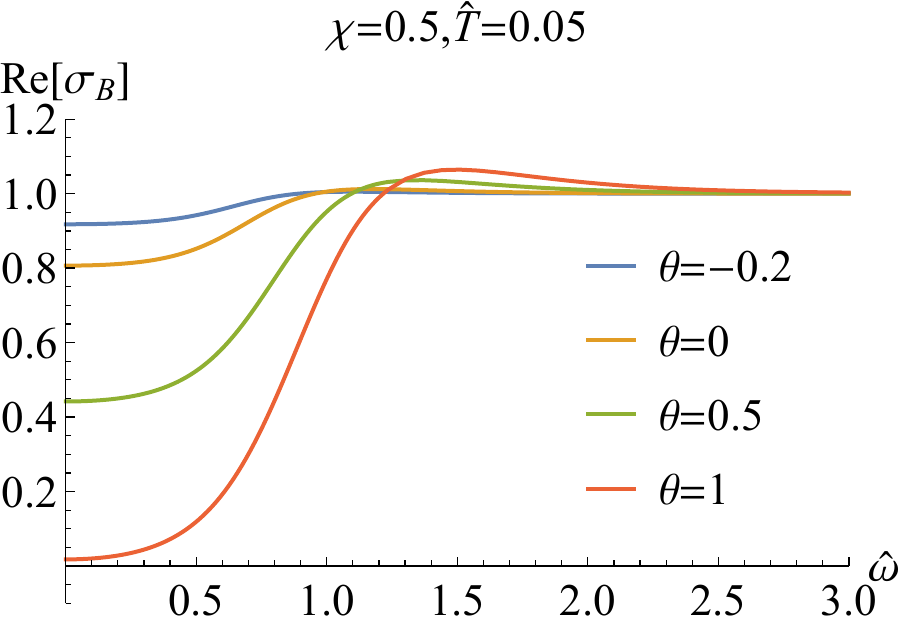}\ \\
\includegraphics[scale=0.8]{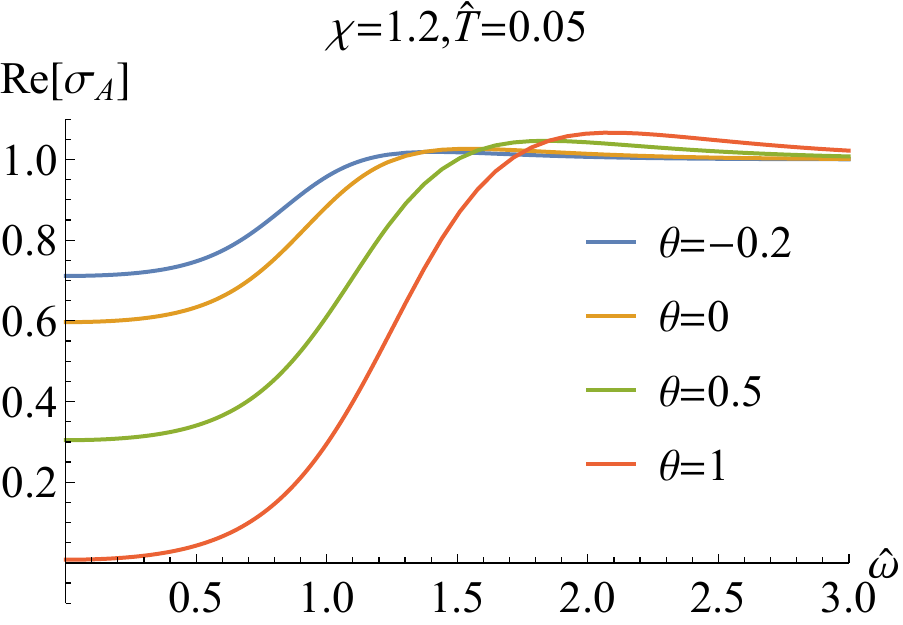}\ \hspace{0.5cm}
\includegraphics[scale=0.8]{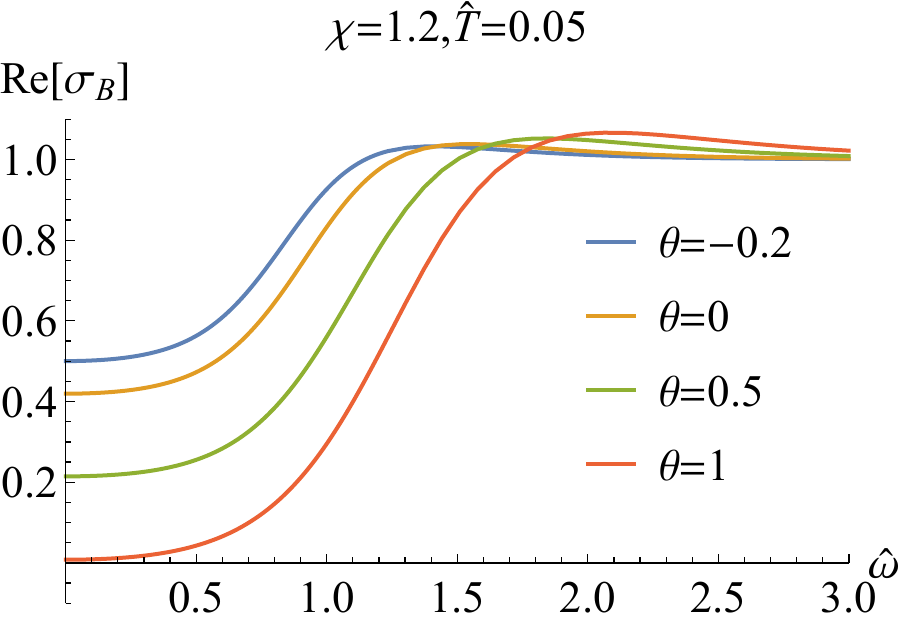}\ \\
\caption{\label{fig_sigmaAB_Chi} The real part of the conductivities $\sigma_A$ (left) and $\sigma_B$ (right) at $\hat{T}=0.05$ for different parameter $\theta$ (the plots above for $\chi=0.5$ and the ones below for $\chi=1.2$).
}}
\end{figure}

Next, we explore the properties of $\sigma_A$ and $\sigma_B$. When $\theta=0$, which has been studied in \cite{Bigazzi:2011ak}, the real part of the conductivities $\sigma_A$/$\sigma_B$ at low frequency exhibits a dip ($\sigma_B$ is frequency independent when $\chi=0$.). With the doping parameter $\chi$ increases, the dip of $Re[\sigma_A]$ becomes shallow but the one of $Re[\sigma_B]$ becomes deepening (see FIG.\ref{fig_sigmaAB_Chiv1} and FIG.\ref{fig_sigmaAB_Chi}). When $\chi$ is fixed, the dip in $Re[\sigma_A]$/$Re[\sigma_B]$ becomes more and more deepening as $\theta$ increases and finally turns into a hard-gap-like when $\theta=1$ is achieved, which is independent of the doping $\chi$. If we further tune $\theta$ larger such that it is beyond the unity, the DC conductivities of $Re[\sigma_A]$/$Re[\sigma_B]$ will be negative, which violate the positive definiteness of the conductivity. Therefore, the positive definiteness of the conductivity imposes a constraint on the coupling parameter $\theta$. Here, we shall constrain $\theta$ in the range of $\theta\leq 1$. Some higher derivative coupling terms also lead to the violation of the positive definiteness of the conductivity \cite{Witczak-Krempa:2013aea,Fu:2017oqa,Gouteraux:2016wxj,Baggioli:2016pia}.

\begin{figure}
\center{
\includegraphics[scale=0.8]{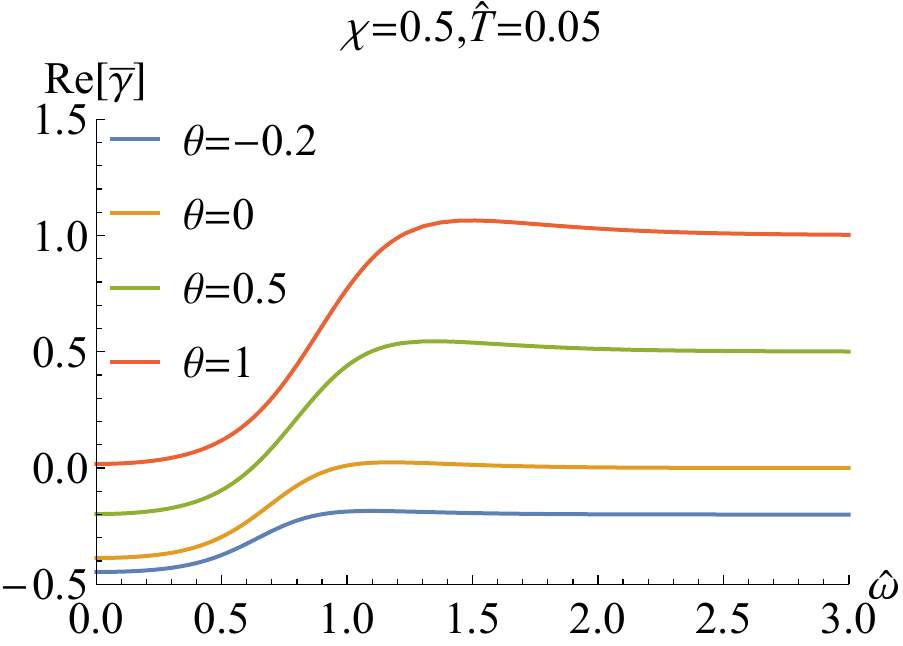}\ \hspace{0.5cm}
\includegraphics[scale=0.8]{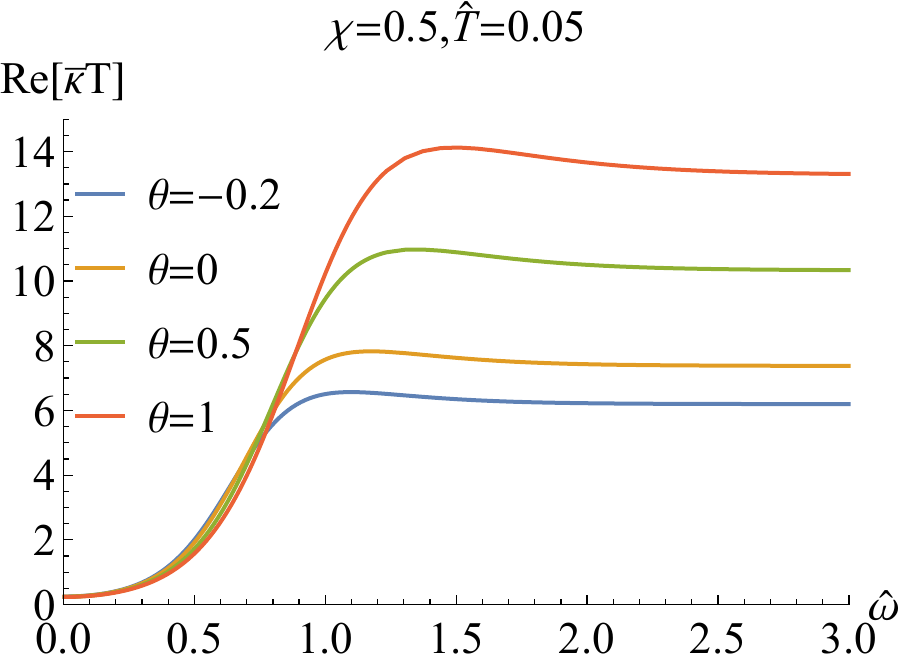}\ \\
\includegraphics[scale=0.8]{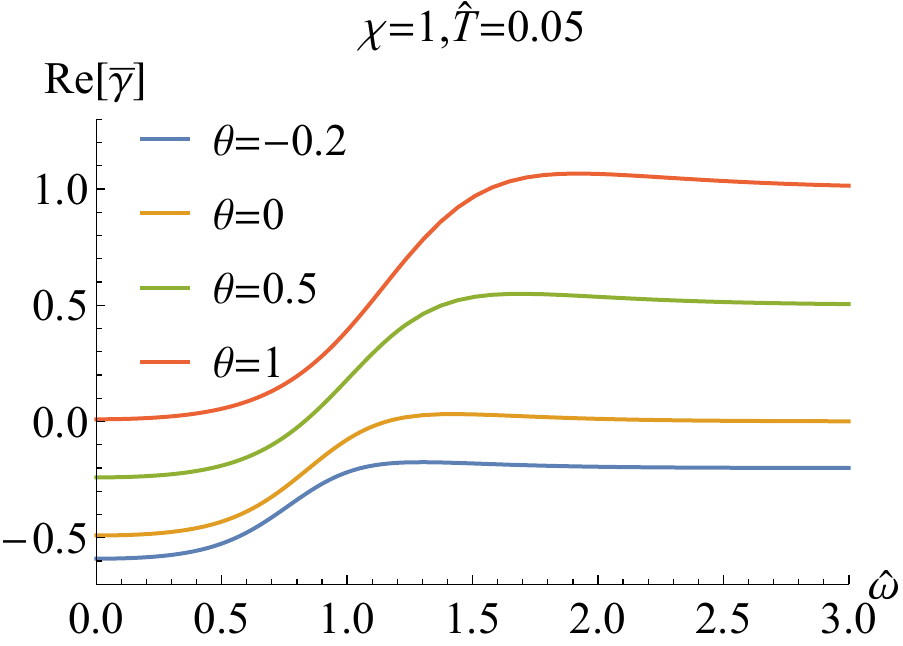}\ \hspace{0.5cm}
\includegraphics[scale=0.8]{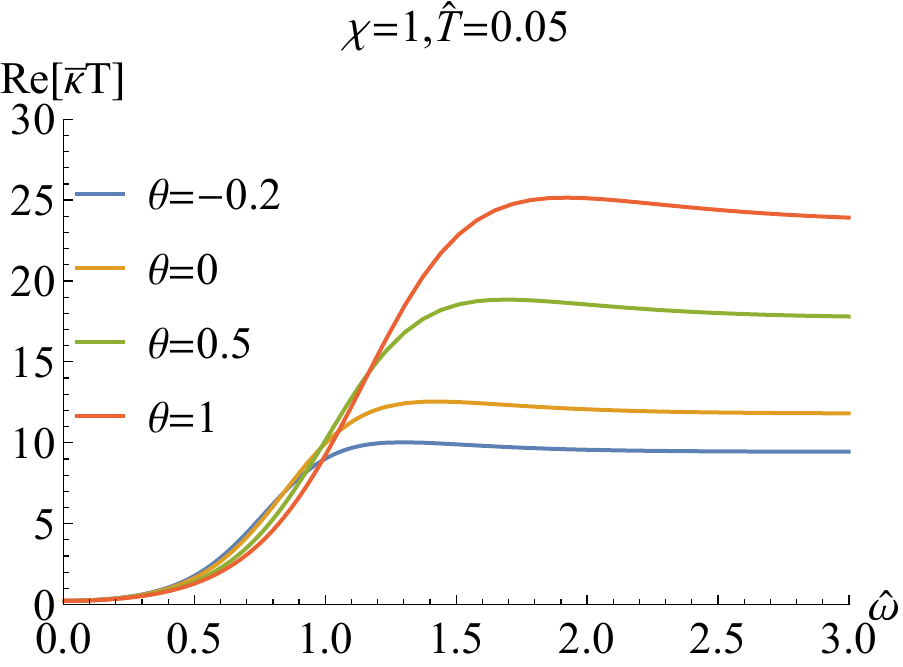}\ \\
\includegraphics[scale=0.8]{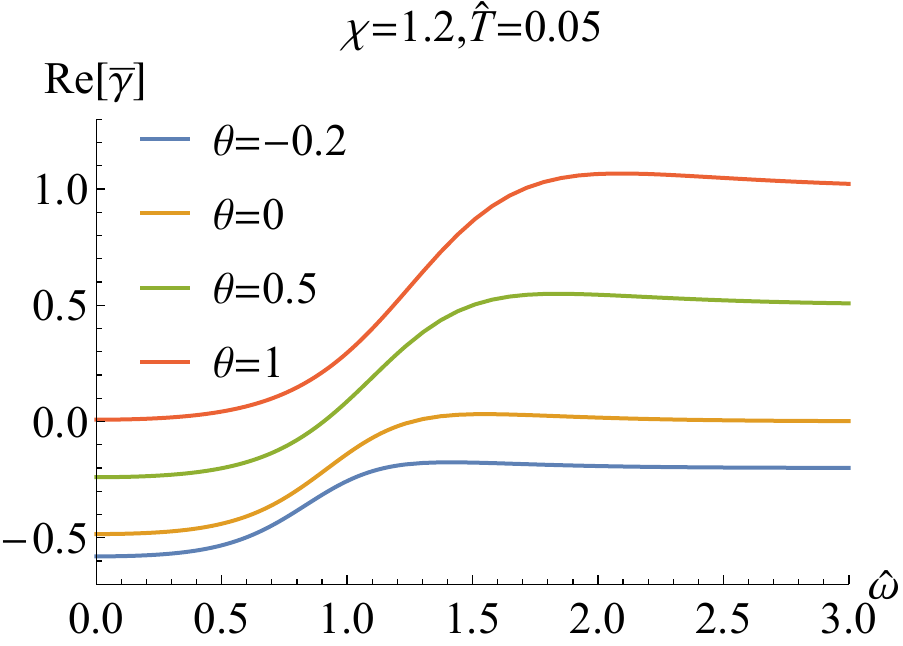}\ \hspace{0.5cm}
\includegraphics[scale=0.8]{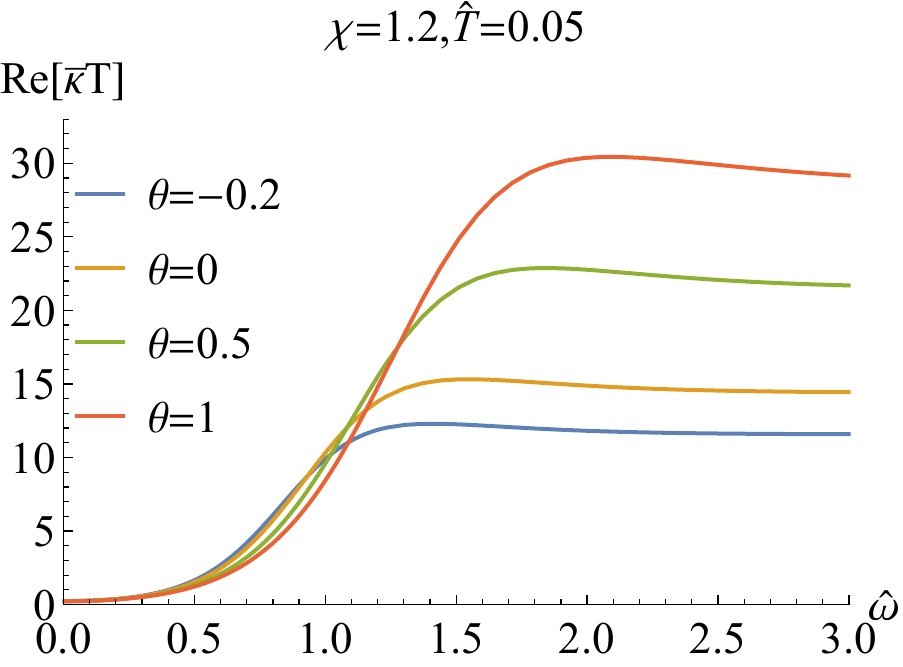}\ \\
\caption{\label{fig_gamma} The real part of $\bar{\gamma}$ and of the thermal conductivity $\bar{\kappa}$ for sample $\chi$ and $\theta$ ($\hat{T}=0.05$).
}}
\end{figure}

Further, we report the real part of $\bar{\gamma}$ and of the thermal conductivity $\bar{\kappa}$ for sample $\chi$ and $\theta$ in FIG.\ref{fig_gamma}. We find that $Re[\bar{\gamma}]$ exhibits a dip at low frequency and approaches a constant at high frequency. At full frequency, $Re[\bar{\gamma}]$ increases as $\theta$ increases.
While $Re[\bar{\kappa}T]$ converges to zero in the limit of $\hat{\omega}\rightarrow 0$ independently of the doping $\chi$ and the coupling $\theta$, which means that the DC thermal conductivity vanishes. As the frequency increases, the thermal conductivities with different $\chi$ and $\theta$ increase and separate out, and then approaches different constant value depending on $\chi$ and $\theta$. With the increase of $\chi$ or $\theta$, this constant value increases.

\begin{figure}
\center{
\includegraphics[scale=0.8]{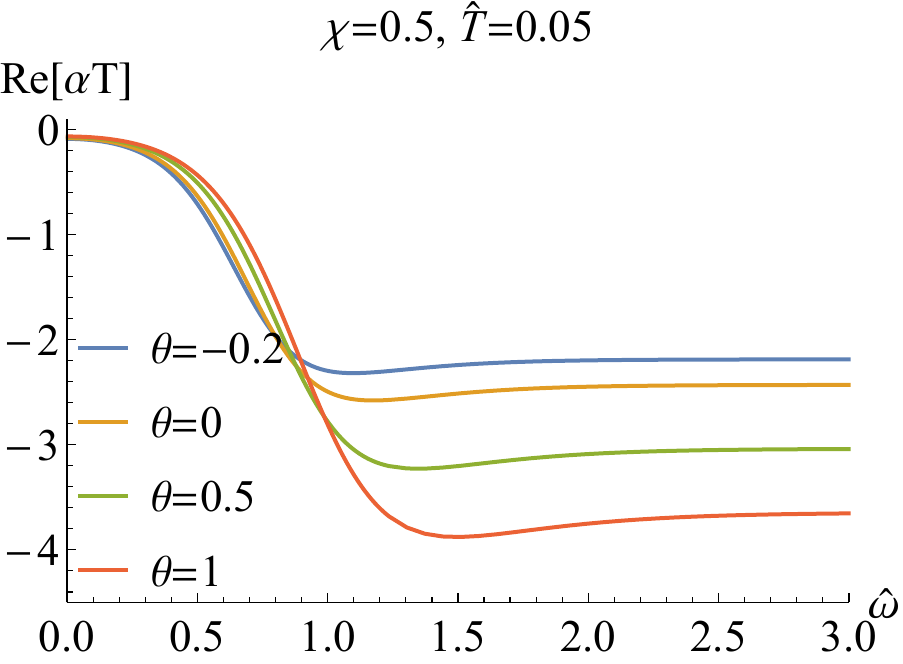}\ \hspace{0.5cm}
\includegraphics[scale=0.8]{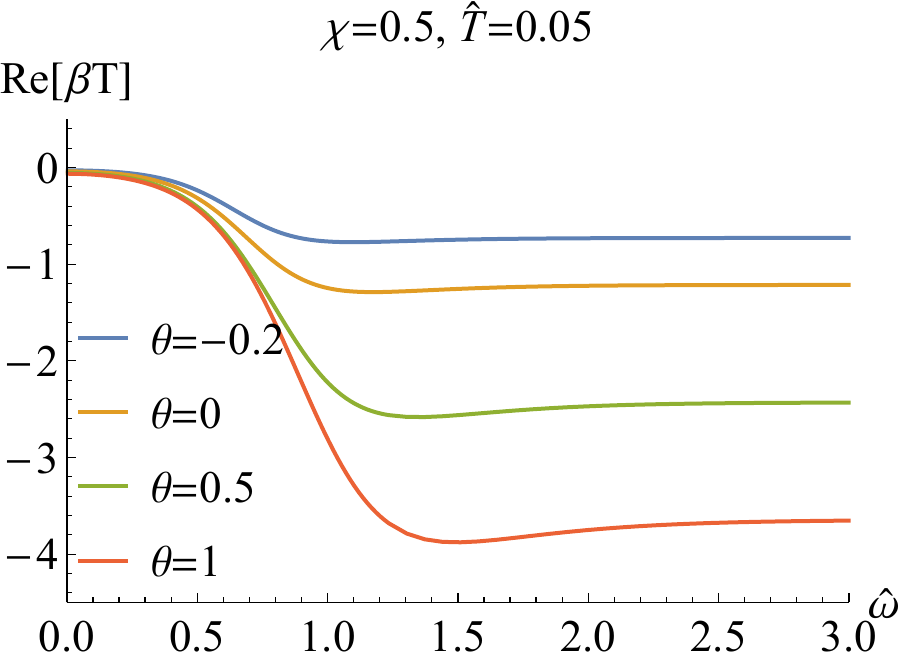}\ \\
\includegraphics[scale=0.8]{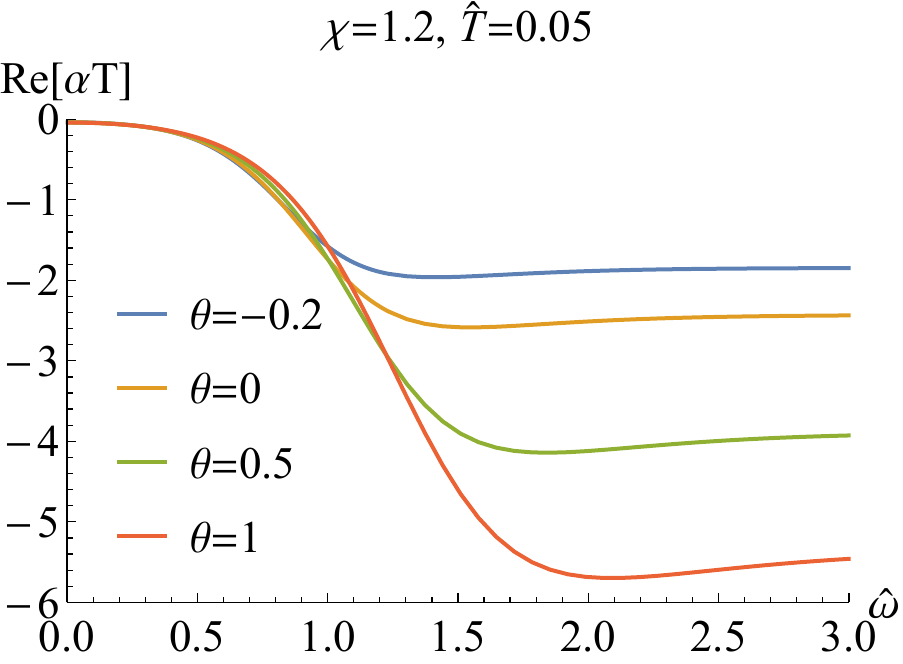}\ \hspace{0.5cm}
\includegraphics[scale=0.8]{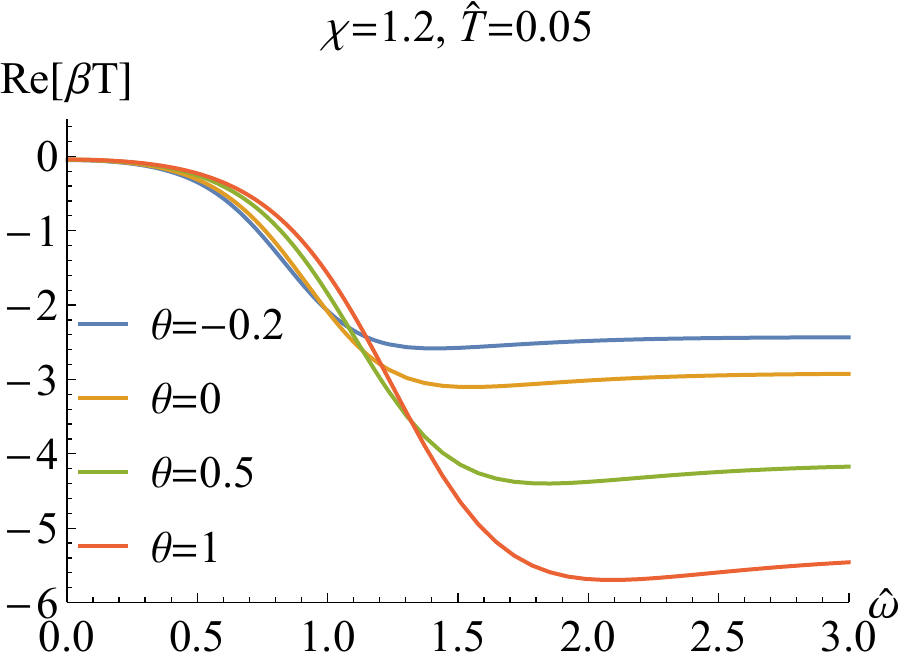}\ \\
\caption{\label{fig_alpha} The real part of the conductivities $\alpha T$ (left) and $\beta T$ (right) for $\chi=0.5$ (above) and $\chi=1.2$ (bottom) at $\hat{T}=0.05$.
}}
\end{figure}

Finally, we also show the real part of the conductivities $\alpha T$ and $\beta T$ for sample $\chi$ and $\theta$ in FIG.\ref{fig_alpha} (also see FIG.\ref{fig_alpha_Chiv1} for $\chi=1$ and different $\theta$). We see that both $\alpha T$ and $\beta T$ are negative and at high frequency, they monotonously decreases with the increase of $\theta$.

\section{Conclusions and discussions}\label{sec-con-dis}

In this paper, we construct a holographic gravity model of two gauge fields with a coupling between them, which corresponds to a two-currents model. An analytical black brane solution is obtained. Also we briefly discuss the thermodynamics. When this coupling is introduced, the expressions of conductivities for holographic two-currents model without coupling studied in \cite{Bigazzi:2011ak} are no longer applicable. By the standard holographic renormalized procedure, we work out the expressions of conductivities with coupling (see Eqs.\eqref{exp-con}). We find that the expressions of conductivities are directly related to the coupling parameter $\theta$. When $\theta=0$, they reduce to that without coupling in \cite{Bigazzi:2011ak}.
The expressions of conductivities for holographic two-currents model with coupling are the main topic of our present paper. Our results are also applicable for the holographic two-currents superconductor model with couple or other extension models.

As an application, then we briefly discuss the properties of the conductivities of this holographic two-currents model with coupling. An interesting property is that as the coupling $\theta$ increases, the dip at low frequency in $Re[\sigma_A]$/$Re[\sigma_B]$ becomes deepening and finally turns into a hard-gap-like when $\theta=1$, which is independent of the doping $\chi$. Some monotonic behaviors of the conductivities are also discussed.

Along this direction, there are lots of works deserving further exploration. For example, we can add a charged complex scalar field to study the superconducting instability and the properties of the conductivities. Also it shall be surely interesting to implement the momentum dissipation into our system with coupling and study the properties of the conductivities.

\begin{acknowledgments}

This work is supported by the Natural Science Foundation of China under
Grant Nos. 11775036, 11905182 and Fok Ying Tung Education Foundation under Grant No. 171006. Guoyang Fu is supported by the Postgraduate Research \& Practice Innovation Program of Jiangsu Province (KYCX20\_2973).
J. P. Wu is also supported by Top Talent Support Program from Yangzhou University.

\end{acknowledgments}

\end{document}